\documentclass[letterpaper]{article} % DO NOT CHANGE THIS

\usepackage{aaai25}  % For camera-ready version

\usepackage{times}  % DO NOT CHANGE THIS
\usepackage{helvet}  % DO NOT CHANGE THIS
\usepackage{courier}  % DO NOT CHANGE THIS
\usepackage[hyphens]{url}  % DO NOT CHANGE THIS
\usepackage{graphicx} % DO NOT CHANGE THIS
\usepackage{natbib}  % DO NOT CHANGE THIS AND DO NOT ADD ANY OPTIONS TO IT
\usepackage{caption} % DO NOT CHANGE THIS AND DO NOT ADD ANY OPTIONS TO IT
\usepackage{algorithm}
\usepackage{algorithmic}
\usepackage{times}
\usepackage{helvet}
\usepackage{courier}
\usepackage{graphicx} % 引入图形处理功能
\usepackage{subcaption} % 子图与子标题

\usepackage{amsmath}
\usepackage{booktabs}
\usepackage{multirow}
\usepackage{diagbox}
\usepackage{makecell}
\usepackage{comment}
\usepackage{bbding}
\usepackage{tabularx}
\usepackage{xcolor}
\usepackage{hhline}

\usepackage{colortbl}

\usepackage{newfloat}
\usepackage{listings}
\DeclareCaptionStyle{ruled}{labelfont=normalfont,labelsep=colon,strut=off} % DO NOT CHANGE THIS
\floatstyle{ruled}
\newfloat{listing}{tb}{lst}{}
\floatname{listing}{Listing}
\title{Seeing Your Speech Style: A Novel Zero-Shot Identity-Disentanglement Face-based Voice Conversion}
\author{
    Yan Rong, Li Liu\thanks{Corresponding Author: avrillliu@hkust-gz.edu.cn.}
}
\affiliations{
    The Hong Kong University of Science and Technology (Guangzhou)\\
}

\usepackage{bibentry}
\begin{document}

\maketitle

\begin{abstract}
\vspace{-0.4em}
\begin{quote}
Face-based Voice Conversion (FVC) is a novel task that leverages facial images to generate the target speaker's voice style. Previous work has two shortcomings: (1) suffering from obtaining facial embeddings that are well-aligned with the speaker's voice identity information, and (2) inadequacy in decoupling content and speaker identity information from the audio input. To address these issues, we present a novel FVC method, \textbf{I}dentity-\textbf{D}isentanglement \textbf{Face}-based \textbf{V}oice \textbf{C}onversion (ID-FaceVC), which overcomes the above two limitations. More precisely, we propose an Identity-Aware Query-based Contrastive Learning (IAQ-CL) module to extract speaker-specific facial features, and a Mutual Information-based Dual Decoupling (MIDD) module to purify content features from audio, ensuring clear and high-quality voice conversion. Besides, unlike prior works, our method can accept either audio or text inputs, offering controllable speech generation with adjustable emotional tone and speed. Extensive experiments demonstrate that ID-FaceVC achieves state-of-the-art performance across various metrics, with qualitative and user study results confirming its effectiveness in naturalness, similarity, and diversity. Project website with audio samples and code can be found at \url{https://id-facevc.github.io}.
\end{quote}
\end{abstract}
\vspace{-1.2em}    
\section{Introduction}
\label{sec:intro}
Voice Conversion (VC) \cite{choi2024dddm,yao2024promptvc} aims to change the speaker identity in speech from a source speaker to that of a target speaker, while preserving the linguistic content. However, audio from the target speaker is not always available in some scenarios (\textit{e.g.}, digital humans, historical figures). Instead, some studies have explored an alternative approach by generating the identity information of \textbf{unseen} speakers' voices from their facial images \cite{mavica2013matching,smith2016concordant}, known as \textbf{Zero-Shot Face-based Voice Conversion (ZS-FVC)}. Recently, this has become a promising research topic with potential applications in various scenarios, such as generating voices that match character appearances in automated film dubbing \cite{cong2024styledubber} and personalized virtual assistants \cite{park2024synthe}.

\begin{figure}[t]
    \vspace{-0.2em}
    \centering
    \includegraphics[width=0.4\textwidth]{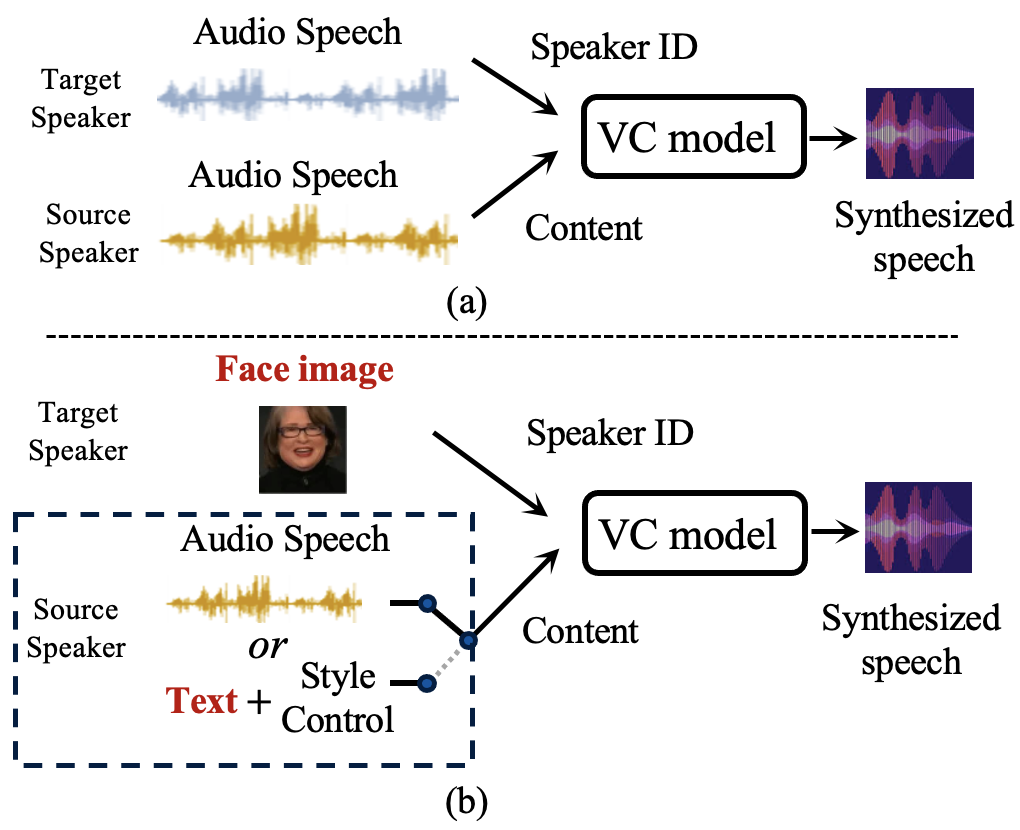}
    \vspace{-0.9em}
    \caption{(a) Traditional voice conversion (VC) paradigm. (b) Our novel ZS-FVC paradigm, which accepts either audio or text as input and allows control over the emotional tone and speed of the generated speech.}
    \label{fig:introduction}
    \vspace{-1.2em}
\end{figure}

In the literature, great progress in this domain has been achieved by prior work \cite{goto2020face2speech,lu2021face,sheng2023face,weng2023zero}. The fundamental challenge is to accurately map identity information between faces and voices. Specifically, this involves (1) \textbf{acquisition of facial embeddings that are well-aligned with the speaker's voice identity}, and (2) \textbf{decoupling of content and speaker identity information from the audio input}.

\textbf{For the first challenge}, the current state-of-the-art (SOTA) work FVMVC \cite{sheng2023face} used FaceNet \cite{schroff2015facenet} to extract general facial features and mapped them through a memory net. Another SOTA work, SP-FaceVC \cite{weng2023zero} averaged all frames to achieve consistent facial embeddings. However, these methods focus on general facial features rather than speaker-specific features, which include substantial non-specific, identity-irrelevant information (\textit{e.g.}, facial expressions, head angles, background). As a result, the models become highly dependent on the training data and lack the ability to locate unique voice characteristics among different speakers, leading to the production of general voices. \textbf{For the second challenge}, FVMVC attempted feature decoupling through a mixed supervision strategy, relying heavily on the quality and scope of the supervision voices. However, in practical scenarios, it is often difficult to acquire adequate and balanced supervision, leading to suboptimal decoupling performance. SP-FaceVC used a low-pass filtering strategy in data pre-processing to eliminate high-frequency elements from audio signals, aiming to reduce style features linked to the speaker identity. Despite its simplicity, this hard filtering approach risks indiscriminately filtering out some key voice details, thereby affecting the naturalness and expressiveness of the synthesized voice and potentially introducing noise and other artifacts.

% To address the above two challenges, we introduce a novel zero-shot \textbf{I}dentity-\textbf{D}isentanglement \textbf{Face}-based \textbf{V}oice \textbf{C}onversion (\textbf{ID-FaceVC)} method. For the first challenge, instead of adopting \textcolor{red}{static encoding methods} that generalize facial features, we design an \textbf{I}dentity-aware \textbf{Q}uery-based \textbf{C}ontrastive \textbf{L}earning (\textbf{IQ-CL}) module to precisely extract the most identity-relevant facial features. Specifically, we employ a set of learnable self-adaptive face prompts to query identity-relevant facial features from a frozen Contrastive Language-Image Pretraining (CLIP) visual encoder \cite{radford2021learning}. \textcolor{red}{The designed Self-Adaptive Face-Prompted QFormer} functions as an information bottleneck, efficiently maps and filters facial features to directly produce speech-relevant facial features, which are then subjected to contrastive learning with identity features extracted from audio. 

To address the above two challenges, we introduce a novel zero-shot \textbf{I}dentity-\textbf{D}isentanglement \textbf{Face}-based \textbf{V}oice \textbf{C}onversion (\textbf{ID-FaceVC)} method. For the first challenge, instead of adopting static encoding methods that generalize facial features, we design an \textbf{I}dentity-\textbf{A}ware \textbf{Q}uery-based \textbf{C}ontrastive \textbf{L}earning (\textbf{IAQ-CL}) module to precisely extract the most identity-relevant facial features. Specifically, we propose a Self-Adaptive Face-Prompted QFormer (SAFPQ), which employs a set of learnable self-adaptive face prompts to query identity-relevant facial features from a frozen Contrastive Language-Image Pretraining (CLIP) visual encoder \cite{radford2021learning}. 
Indeed, the SAFPQ functions as an information bottleneck, efficiently filters and maps facial features to produce speech-relevant facial features, which are then subjected to contrastive learning with identity features extracted from audio.

\begin{table}[t] \small
\vspace{-0.1em}
\caption{Comparison of input ways and controllability with previous studies.}
\vspace{-1.3em}
\centering
\setlength{\tabcolsep}{6pt}
\resizebox{0.48\textwidth}{!}{
\tabcolsep 0.14in
\begin{tabular}{c|c|c}
\bottomrule[1.3pt]
\textbf{Methods} & \textbf{Input Modalities} & \textbf{Controllability}\\
\midrule[1.1pt]
Face2Speech \cite{goto2020face2speech} & Text & \XSolidBrush \\
FaceVC \cite{lu2021face} & Audio & \XSolidBrush \\
FVMVC \cite{sheng2023face}& Audio & \XSolidBrush \\
SP-FaceVC \cite{weng2023zero} & Audio & \XSolidBrush \\
FaceTTS \cite{lee2023imaginary} & Text & \XSolidBrush \\
\midrule[1.1pt]
\textbf{Ours (ID-FaceVC)} &Audio / Text& \Checkmark\\
\bottomrule[1.3pt]
\end{tabular}
}
\label{table:intro}
\vspace{-1.8em}
\end{table}

For the second challenge, rather than using implicit supervision or hard filters, we design a novel \textbf{M}utual \textbf{I}nformation-based \textbf{D}ual \textbf{D}ecoupling \textbf{(MIDD)} module to purify the extracted content features. This module decomposes speech into subspaces representing different attributes and minimizes the overlapping information between speaker identity and content features through Mutual Information (MI) constraints. Additionally, inspired by \cite{peng2023emotalk,park2024synthe}, we implement the fine-grained speaker identity supervision to fully leverage speaker identity information, compelling the model to learn the subtle distinctions between different speakers and preventing model collapse.

In addition, previous approaches that employed the target speaker's voice as input \cite{lu2021face,sheng2023face,weng2023zero}, as depicted in Table \ref{table:intro}, suffer from limitations in practical applications due to the occasional unavailability of the reference audio. Some existing works have utilized text as the input for speech generation \cite{goto2020face2speech,lee2023imaginary}, but their outputs lack the flexibility to manipulate speech style and often produce speech in a ``machine" manner. In this work, we first incorporate text as an alternative modality during the inference stage and introduce a style-controllable strategy that allows for control over the emotion and speed of the generated speech, thereby enabling the generation of natural, rhythmical, and controllable speech from text.

In summary, the main contributions of this work are as follows.
\begin{itemize}
\item A novel paradigm named ID-FaceVC is proposed for zero-shot face-based voice conversion that can accept either audio or text as input, allowing control over the emotional tone and speed of the generated speech. To the best of our knowledge, this is the first attempt to explore dual-input controllable face-based voice conversion.
\item We design an IAQ-CL module, containing a new Self-Adaptive Face-Prompted QFormer to query facial features most relevant to speaker identity and forces the model to learn the subtle differences between speakers.
\item We propose an effective mutual information-based MIDD module to completely decouple content and speaker identity from audio features. 
\item Extensive experimental results demonstrate that our method achieves SOTA performance across multiple metrics. Qualitative and user study results further validate the effectiveness of the proposed model in terms of naturalness, similarity, and diversity.
\end{itemize}

\begin{comment}
\begin{figure*}[t]
    \vspace{-0.4em}
    \centering
    \hspace{0.25em}
    \begin{subfigure}[b]{0.98\columnwidth}
        \centering
        \includegraphics[width=\textwidth]{sup/network_train.png}
        \vspace{-1.4em}
        \caption{The training stage}
        \vspace{-0.7em}
        \label{fig:train}
    \end{subfigure}
    \hfill
    \begin{subfigure}[b]{1.04\columnwidth}
        \centering
        \includegraphics[width=\textwidth]{sup/network_infer.png}
        \vspace{-1.4em}
        \caption{The inference stage}
        \vspace{-0.7em}
        \label{fig:inference}
    \end{subfigure}
    \hspace{0.25em}
    \caption{Overview of our proposed ID-FaceVC, where SAFPQ represents Self-Adaptive Face-Prompted QFormer. (a) During the training process, paired face-voice data serve as input. (b) In the inference stage, the content of the generated speech can be controlled via text or voice inputs. The model can generate speech that matches the style of the target speaker's facial image.}
    \label{fig:overview}
    \vspace{-1.3em}
\end{figure*}
\end{comment}

\begin{figure*}[th]
    \vspace{-0.2em}
    \centering
    \includegraphics[width=0.99\textwidth]{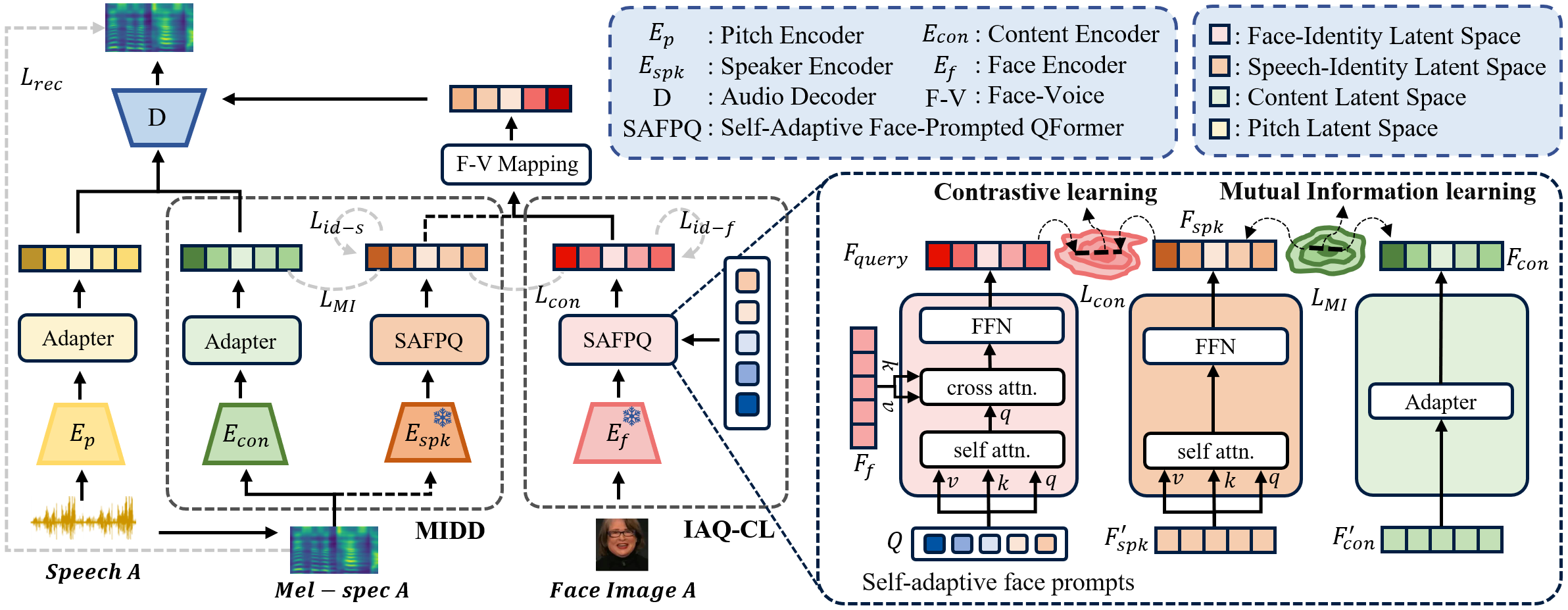}
    \vspace{-0.7em}
    \caption{Overview of the proposed ID-FaceVC. The Adapter is a Feed Forward Network used to adjust vector dimensions. The embeddings $F_{f}$, $F_{spk}^{’}$, $F_{con}^{’}$ correspond to the face, speaker, and content features extracted by $E_f$, $E_{spk}$, $E_{con}$, respectively.}
    \label{fig:overview}
    \vspace{-1.2em}
\end{figure*}
\section{Related Work}
\label{sec:relatedwork}

\subsection{Evidence of Face-Voice Correlation}
Facial and vocal characteristics are closely linked to individual identity. Studies have demonstrated a natural synergy between these features, collectively providing concordant source identity information \cite{smith2016concordant}. Features of a voice can be inferred from facial structures \cite{krauss2002inferring,mavica2013matching}. For example, vocal pitch and intonation may be associated with facial features like jaw width and eyebrow density, which together create a distinct identity signature. Recently, several studies have exploited the strong similarity between voice and face for novel applications, such as reconstructing a speaker’s face from their voice \cite{wang2023realistic,oh2019speech2face,wen2019face,duarte2019wav2pix}. Our research explores the inverse of this process, generating diverse vocal styles from various facial images.

\subsection{Face-based Voice Conversion}
Prior research has validated the potential for synthesizing speech from facial features. Face2Speech \cite{goto2020face2speech} pioneered this field with a three-stage training strategy and a supervised generalized end-to-end loss to generate speech that reflects speaker facial characteristics. Building on this foundation, subsequent works proposed more adaptable loss functions \cite{wang2022residual} and more sophisticated network designs \cite{lee2023imaginary} to enhance the quality of the synthesized speech. These methodologies typically employ text as the input to avoid entanglement issues. FaceVC \cite{lu2021face} developed a three-stage model that leverages a bottleneck adjustment strategy and a straightforward MSE loss to extract necessary content embeddings from audio. However, this model struggles to capture the complex mappings between speech and facial domains, often defaulting to predicting an ``average voice" across variations, making it unsuitable for zero-shot applications. The most advanced approaches in this field, FVMVC \cite{sheng2023face} and SP-FaceVC \cite{weng2023zero}, improved upon FaceVC through memory-based feature mapping and rigorous data preprocessing.

Nevertheless, these methods still have considerable potential for improvement in achieving well-aligned facial embeddings with speech and effectively decoupling content from speaker identity in audio features.

%\subsection{Text-based Voice cloning}

\section{Our Method}
\label{sec:Method}

Our proposed ID-FaceVC employs an end-to-end training approach. It comprises three main components: ID-Aware Query-based Contrastive Learning module, Mutual Information-based Dual Decoupling module, and Alternative Text-Input with Style Control module.
%For input facial images, we construct the IDQ-CL module, designed to extract features most relevant to the speaker's identity. In handling paired speech features, we designed the MIDD module that employs multiple latent spaces to disentangle various attributes and minimize the overlap of speaker identity and content information via mutual information. Additionally, speaker-identity supervision is applied to train the model to learn key identity-specific features between different speakers. During the inference phase, we introduce text as an additional control method for speech content, and through a style controllable strategy, we ensure that the speech generated from text is natural, rhythmic, and adjustable.

\subsection{ID-Aware Query-based Contrastive Learning}
We design the IAQ-CL module to extract facial features that are well-aligned with the speaker’s voice identity. This module includes Self-Adaptive Face-Prompted QFormer and face-related speaker identity supervision.

\begin{comment}
\begin{figure}[t]
    \vspace{-0.4em}
    \centering
    \includegraphics[width=0.45\textwidth]{sup/Qformer.png}
    \vspace{-0.3em}
    \caption{The architecture of the self-adaptive face-prompted QFormer. The embeddings $F_{f}$, $F_{spk}^{’}$, $F_{con}^{’}$ correspond to the face, speaker, and content features extracted by $E_f$, $E_{spk}$, $E_{con}$, respectively.}
    \label{fig:qformer}
    \vspace{-1.2em}
\end{figure}
\end{comment}

\subsubsection{Self-Adaptive Face-Prompted QFormer.} 
Considering the inherent limitation of CNN-based architectures in handling the diversity of facial features, we instead employ a frozen CLIP visual encoder \cite{radford2021learning} to extract features from facial images. For the frame-wise visual embeddings extracted by the CLIP visual encoder, we compute the arithmetic mean to obtain average frame-level facial features, rather than randomly selecting a single frame as the facial embedding, to reduce potential sampling bias.

Due to the high-dimensional and redundant nature of CLIP visual features, facial embeddings contain abundant information, including facial expressions, head poses, and backgrounds, with only a small portion related to the speaker's style. Therefore, we propose the SAFPQ to filter the most speech-relevant features, as illustrated in Figure \ref{fig:overview}. Unlike the vanilla Query Transformer (QFormer) \cite{li2023blip}, our model better integrates identity information from both face and voice domains, resulting in a more cohesive representation. The SAFPQ functions as an information bottleneck, filtering out redundant facial features while emphasizing those crucial for speech. In the inference stage, the self-adaptive face prompts retrieves identity-relevant facial features from input facial embeddings, facilitating the prediction of the speaker's style from unseen facial images. 

To be specific, we initialize a set of learnable self-adaptive face prompts. The most informative prompts are highlighted through a self-attention mechanism that integrates the information from a global perspective. Subsequently, the face prompts interact with facial embeddings via cross-attention to retrieve features relevant to the identity information. Finally, a fully connected layer fuses these retrieved features. The process are defined as follows:
\vspace{-0.4em}
\begin{equation}
A_{self}=\operatorname{softmax}\left(\frac{\mathbf{Q} W_{q}^{\text {self }}\left(\mathbf{Q} W_{k}^{\text {self }}\right)^{T}}{\sqrt{d_{k}}}\right) \mathbf{Q} W_{v}^{\text {self }} ,
\end{equation}
\begin{equation}\small
    A_{cross}= \operatorname{softmax}\left(\frac{A_{self} W_{q}^{\text {cross}}\left(F_f W_{k}^{\text {cross}}\right)^{T}}{\sqrt{d_{k}}}\right) F_f W_{v}^\text {cross} ,
\end{equation}
\begin{equation} \small
    F_{query}=\operatorname{FFN}(A_{cross}) ,
\end{equation}
where $W_{q}^{\text{self}}$, $W_{k}^{\text{self}}$, $W_{v}^{\text{self}}$ are the learnable weights for the self-attention, and $W_{q}^{\text{cross}}$, $W_{k}^{\text{cross}}$, $W_{v}^{\text{cross}}$ are the learnable weights for the cross-attention. $Q$ is the self-adaptive face prompts, $d_k$ is the dimension of the key, $F_f$ is the facial embedding extracted by the CLIP, and $F_{query}$ represents the final queried facial features.

Recall that our objective is to extract features highly relevant to the speaker's identity, ensuring that the retrieved facial embeddings closely match the style features in speech. We employ contrastive learning to measure the distance between these two features, thereby optimizing the self-adaptive face prompts. This encourages the speech style and facial embeddings from the same speaker to be as similar as possible, while those from different speakers are distinctly separated. The formulation for this process is as follows:
\vspace{-0.2em}
\begin{equation} \small
\label{Eq:con}
L_{con}=-\frac{1}{N} \sum_{i=1}^{N} \sum_{j=1}^{N} y_{i, j} \log \left(\frac{\exp (\operatorname{sim}(i, j)/\tau)}{\sum_{k=1}^{N} \exp (\operatorname{sim}(i, k)/\tau)}\right),
\end{equation}
where $N$ is the number of samples in a batch, $\tau$ is a temperature hyperparameter, $i$ is a fixed index for facial embeddings, $j$ and $k$ are indices for speech embeddings, and $y_{i, j}$ is an indicator function. If samples $i$ and $j$ belong to the same speaker, then $y_{i, j} = 1$; otherwise, $y_{i, j} = 0$.

\begin{figure}[t]
    \vspace{-0.4em}
    \centering
    \includegraphics[width=0.49\textwidth]{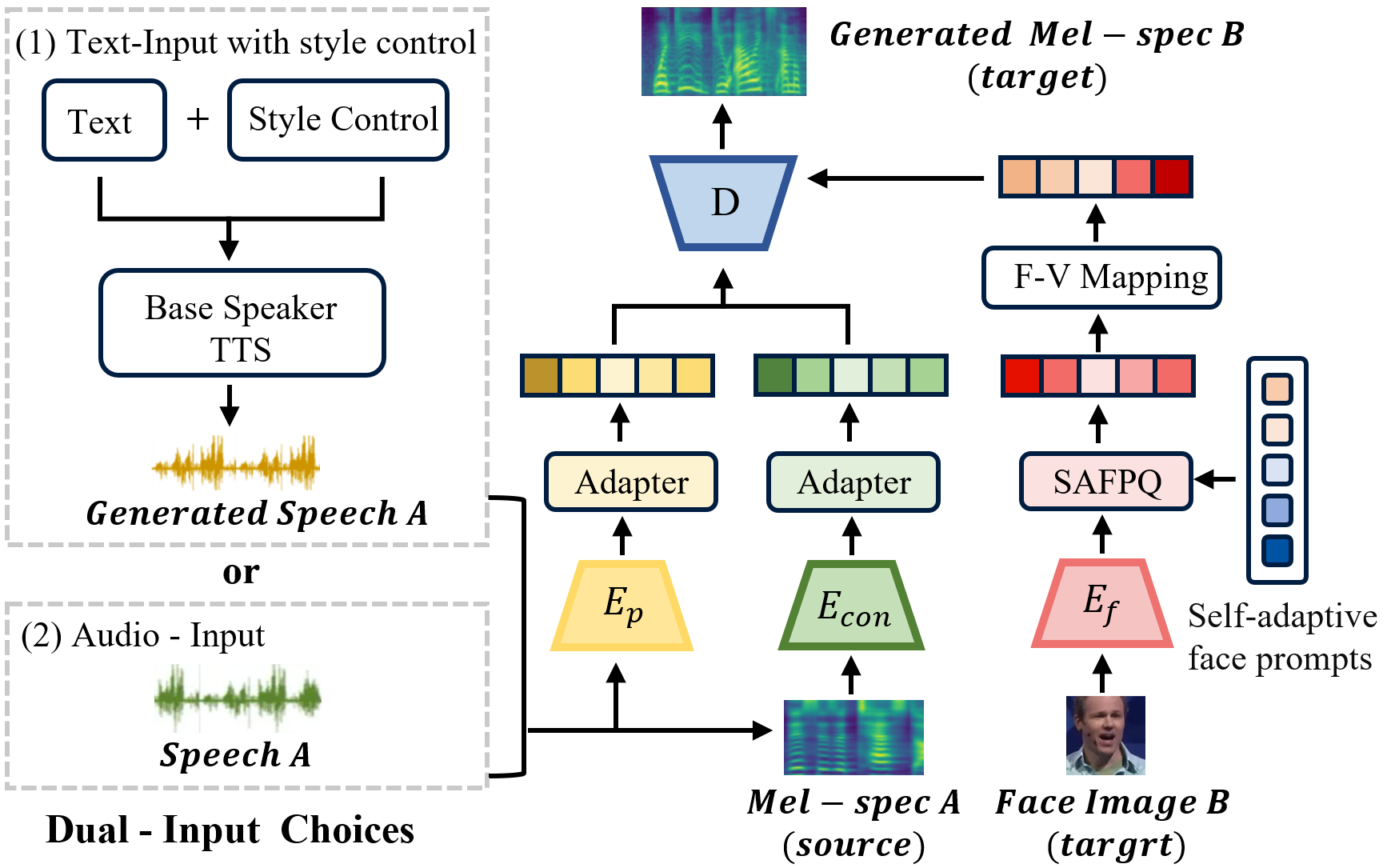}
    \vspace{-1.2em}
    \caption{The inference stage of ID-FaceVC. Text is introduced as an alternative modality to produce natural, rhythmic, and controllable speech.}
    \label{fig:inference}
    \vspace{-1.2em}
\end{figure}

\subsubsection{Face-related Speaker-identity Supervision.} To distinguish key facial features between different speakers, inspired by \cite{peng2023emotalk}, we design a fine-grained speaker identity supervision mechanism to enhance our model's capability. This supervision ensures that facial features should maintain consistency for the same speaker while exhibiting distinctiveness between different speakers. The formulation can be expressed as:
\vspace{-0.5em}
\begin{equation} \small
\label{Eq:idf}
L_{i d-f}=-\frac{1}{N} \sum_{n=1}^{N} \sum_{i=1}^{C} t_{n i} \cdot \log \left(p_{n i}\right),
\end{equation}
where $C$ represents the number of distinct speakers and $t_{ni}$ denotes the one-hot encoded target label for the $n$-th sample (where $t_{ni}=1$ if the sample belongs to the $i$-th speaker, otherwise $t_{ni}=0$). $p_{n i}$ is the softmax probabilities that the $n$-th sample’s $F_{query}$ belong to the $i$-th speaker. 

During the inference, ID-FaceVC is able to generate consistent style speech for different facial images of the same speaker and diverse style speech for different speakers.

\subsection{Mutual Information-based Dual Decoupling}
We propose the MIDD module to achieve precise representation of different disentangled latent spaces and purification of speech content information. It includes Disentangled Latent Space and Mutual Information-based Decoupling.

\subsubsection{Disentangled Latent Space.}
The core of achieving robust content representation is the removal of non-content-related features. A natural idea is to decompose speech into distinct subspaces that represent various attributes. Thus, we use two different encoders to separately extract a compact speaker style code $F_{spk}$ and a continuous content code $F_{con}$ from the mel spectrogram. Specifically, to fully leverage the powerful representational capabilities of large models, we employ the Contrastive Language-Audio Pretraining (CLAP) \cite{wu2023large} audio encoder as our speaker encoder. As depicted in Figure \ref{fig:overview}, the features obtained from the CLAP are processed through SAFPQ, following the same procedure as facial embedding handling but without the cross-attention module. For the content encoder, we adopt vector quantization \cite{bitton2020vector} and contrastive predictive coding \cite{oord2018representation} techniques, commonly used in voice conversion tasks, to extract the content embeddings $F_{con}$.

\subsubsection{Mutual Information-based Decoupling.}
Given the diversity of speech styles, merely constructing two separate latent spaces may not ensure sufficient feature decoupling. Previous decoupling methods, such as inter-speaker supervision \cite{schroff2015facenet}, still result in certain overlaps among features. To address this issue, we utilize MI, which can measure the overall dependency between variables and capture both linear and non-linear relationships \cite{veyrat2009mutual}, as a metric to evaluate the correlation between speaker embeddings and content embeddings extracted from speech. However, due to the high dimensionality and unknown distributions of the variables, directly calculating probability distributions in MI is impractical. To solve this, we employ a variational upper bound technique \cite{cheng2020club}, to establish parameterized conditional distributions, which aids in controlling the minimization process by estimating the upper bound of MI:
\vspace{-0.2em}
\begin{equation} \small
\label{Eq:MI}
L_{MI}=\frac{1}{N^{2}} \sum_{i=1}^{N} \sum_{j=1}^{N} \log \frac{q\left(F_{\mathrm{con}, i} \mid F_{\mathrm{spk}, i}\right)}{q\left(F_{\mathrm{con}, j} \mid F_{\mathrm{spk}, i}\right)},
\end{equation}
where $F_{\text{spk}, i}$ denotes the speaker style embeddings for the $i$-th sample, $F_{\text{con}, i}$ and $F_{\text {con}, j}$ represent the content embeddings for the $i$-th and $j$-th samples, respectively. 

By minimizing the overlap information between the extracted style features and content features, we successfully establish a speaker style space related to identity and a content space associated with semantics.

%This metric aims to purify the content features derived from speech. MI is calculated using this formula: 
% \vspace{-0.5em}
% \begin{equation} \small
% I(X; Y)=E_{p(x, y)}\left[\log \frac{p(x, y)}{p(x) p(y)}\right].
% \end{equation}

% \subsubsection{Speech-related Speaker-identity Supervision.}
% Similar to \textcolor{red}{facial features}, we apply speaker-identity supervision to the style features extracted from speech:
% \vspace{-0.5em}
% \begin{equation} \small
% \label{Eq:ids}
% L_{i d-s}=-\frac{1}{N} \sum_{n=1}^{N} \sum_{i=1}^{C} t_{n i} \cdot \log \left(p_{n i}^{(s)}\right).
% \end{equation}
% where $p_{n i}^{(s)}$ represents the softmax probability that the speaker feature $F_{spk}$ of the $n$-th sample belongs to the $i$-th speaker, with other variables defined consistently with Eq. (\ref{Eq:idf}). 

In addition, similar to Eq. (\ref{Eq:idf}), we apply speech-related speaker-identity supervision $L_{id-s}$ to the style features extracted from speech. In this context, $p_{n i}$ in represents the softmax probability that the speaker feature $F_{spk}$ of the $n$-th sample belongs to the $i$-th speaker, while other variables remain consistent with Eq. (\ref{Eq:idf}). The joint speaker-identity supervision for both facial and speech features enforces the model to recognize the consistency within the same speaker's identity and the diversity across different speakers' identities. These two loss functions prevent the generation of overly similar outputs, thereby protecting against mode collapse.

\begin{table*}[thb]
\centering
\caption{Comparison with SOTA methods. Best performances are highlighted in bold, while second-best are underlined.}
\vspace{-0.9em}
\label{SOTA_compare}
%\resizebox{73mm}{25mm}{
\resizebox{\textwidth}{!}{
    \begin{tabular}{cccccccc}
    \toprule
    \multirow{2}{*}{Input Modalities}&\multirow{2}{*}{Methods} &\multicolumn{3}{c}{Naturalness} &\multicolumn{1}{c}{Similarity} &\multicolumn{2}{c}{Consistency \& Diversity}\\
    \cmidrule(r){3-5} \cmidrule(r){6-6} \cmidrule(r){7-8} 
    && UTMOS $\uparrow$ & WER $\downarrow$& CER $\downarrow$& SECS $\uparrow$  & SEC $\uparrow$ & SED $\downarrow$\\
    \midrule
    \multirow{4}{*}{Audio Input}&FaceVC \cite{lu2021face}& 2.155& \underline{16.67\%}&\underline{10.79\%} &0.702  & 0.986 &0.965 \\
    &SP-FaceVC \cite{weng2023zero}& 1.831&29.17\% &19.67\% &\textbf{0.723} &\textbf{0.988}&0.912  \\
    &FVMVC \cite{sheng2023face}&\underline{3.023}  &21.79\% & 15.02\%& 0.678 & 0.987 &\underline{0.835} \\
    &Ours (ID-FaceVC)& \textbf{3.286} & \textbf{12.11\%}&\textbf{7.86\%} &\underline{0.713} &  \textbf{0.988} & \textbf{0.832} \\
    \midrule
    \multirow{2}{*}{Text Input}&FaceTTS \cite{lee2023imaginary} &2.102 & 14.31\%& 8.70\%&  0.701&0.987&0.912 \\
    &Ours (ID-FaceVC)&\textbf{3.454} & \textbf{5.02\%}& \textbf{2.69\%}& \textbf{0.704}& \textbf{0.989 }& \textbf{0.844}\\
    \bottomrule
    \end{tabular}
}
\end{table*}

\begin{figure*}[th]
    \vspace{-0.2em}
    \centering
    \includegraphics[width=0.95\textwidth]{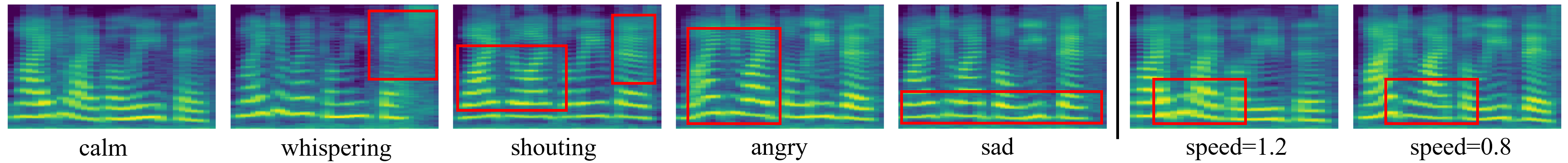}
    \vspace{-0.8em}
    \caption{Mel-spectrogram visualizations of voices generated from text inputs under different emotions and speeds. The red boxes highlight areas with significant changes 
    compared with the ``calm" state.}
    \label{fig:melcontrol}
    \vspace{-1.2em}
\end{figure*}

\subsection{Alternative Text-Input with Style Control}
In addition to audio inputs, text serves as a more flexible modality in practical applications because it does not require prior recording of a source speaker's speech. In this work, as illustrated in Figure \ref{fig:inference}, we introduce text as an alternative option to specify the content of generated audio, which broadens the applicability and accessibility of our framework.

%\subsubsection{Style Control for Text Input.}
The transition from text to audio often results in monotonous narrations due to the absence of references for emotion, accent, and rhythm. To address this issue, inspired by the OpenVoice \cite{qin2023openvoice}, we develop a style control strategy that uses the base speaker TTS as a bridge to generate an intermediate single-speaker audio. This audio can be flexibly manipulated in terms of speed and emotion through style control parameters. The choice of base speaker TTS is flexible, allowing for either a single-speaker or multi-speaker TTS, as the timbre produced by the TTS is not our focus. In this task, we select the VITS \cite{kim2021conditional} model as the base speaker TTS, which accepts both text and style control inputs. The audio generated through this process then serves as the source speaker audio input into our network, where a content encoder extracts the speech content, and a pitch encoder captures the pitch information. Together with the speaker style information inferred from unseen speaker facial images, we generate the final audio output.

In the inference, when using text as input, the flexible control of speech and the injection of timbre inference are separated, allowing for a straightforward and training-free implementation of face-based controllable voice conversion.

\subsection{Training Loss}
We utilize L2 loss to evaluate the quality of the reconstructed mel spectrograms. The formula for this is as follows:
\vspace{-0.1em}
\begin{equation} \small
L_{r e c}=\|Mel-\hat{Mel}\|_{2}^{2} ,
\end{equation}
\vspace{-0.1em}where $Mel$ and $\hat{Mel}$ represent the Mel spectrogram input to the network and the Mel spectrogram reconstructed by the model, respectively. Additionally, we follow the training setup described in FVMVC \cite{sheng2023face}, incorporating both inter-speaker supervision loss and face-voice mapping loss, collectively referred to as $L_{F}$. 

The total training loss is defined as follows:
\vspace{-0.3em}
\begin{equation} \small
\mathcal{L}= L_{rec}+\lambda_{1} L_{con}+\lambda_{2} L_{MI}+\lambda_{3} L_{i d-f}+\lambda_{4} L_{i d-s}+\lambda_{5} L_{F},
\label{Eq:loss}
\end{equation}
where $\lambda_{1}$ is the weight of $L_{con}$ (in Eq. (\ref{Eq:con})), $\lambda_{2}$ is the weight of $L_{MI}$ (in Eq. (\ref{Eq:MI})), $\lambda_{3}$ is the weight of $L_{id-f}$ (in Eq. (\ref{Eq:idf})), $\lambda_{4}$ is the weight of $L_{id-s}$, and $\lambda_{5}$ is the weight of $L_{F}$.

\section{Experiment and Result}
\label{sec:Experiment}

\subsection{Experimental Setup}
\subsubsection{Datasets.} To the best of our knowledge, current ZS-FVC methods utilized the LRS3 \cite{afouras2018lrs3} dataset, which comprises over 400 hours of TED talks collected from YouTube, for training. For a fair comparison, we follow the same dataset setup. More precisely, we selected the paired data from the top 200 speakers by video count, resulting in 11,430 videos for training and 5,173 videos for validation. For testing, we randomly selected 16 previously unseen speakers, including 8 target speakers (4 male, 4 female) and 8 source speakers (4 male, 4 female).

\subsubsection{Implementation Details.} We employ the MTCNN \cite{zhang2016joint} to detect and align faces in each video frame. Facial features are extracted using the ViT-B/32 from CLIP, with outputs from the penultimate layer utilized to enhance generalization over the final layer. Audio is extracted from video clips via FFmpeg \cite{yamamoto2020parallel}, and the HTSAT-base from CLAP serves as the speaker feature extractor. Training is conducted on a single Nvidia-A800 GPU with a batch size of 256 for 2000 epochs. F-V mapping is a memory-based feature mapping module, following the setup of FVMVC \cite{sheng2023face}. For the vocoder, we utilize a pretrained ParallelWaveGAN \cite{yamamoto2020parallel}. Loss weights specified in Eq. (\ref{Eq:loss}) are set at $\lambda_{1}=0.1$, $\lambda_{2}=0.01$, $\lambda_{3}=0.1$, $\lambda_{4}=0.1$, and $\lambda_{5}=1$.

\begin{figure*}[th]
    \vspace{-0.2em}
    \centering
    \includegraphics[width=0.76\textwidth]{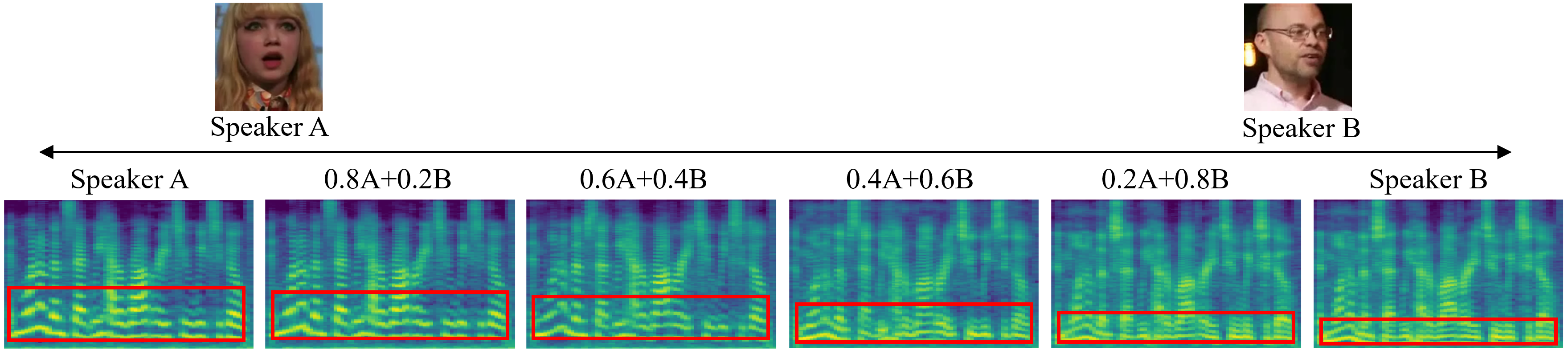}
    \vspace{-0.8em}
    \caption{The mel-spectrograms of voices generated by mixed facial embeddings. From left to right, as the weight of male facial embeddings increases, the voice characteristics gradually shift from female to male, and the fundamental frequency decreases.}
    \label{fig:mix_faceemb}
    \vspace{-1.2em}
\end{figure*}

\begin{figure}[t]
    \vspace{-0.2em}
    \centering
    \includegraphics[width=0.35\textwidth]{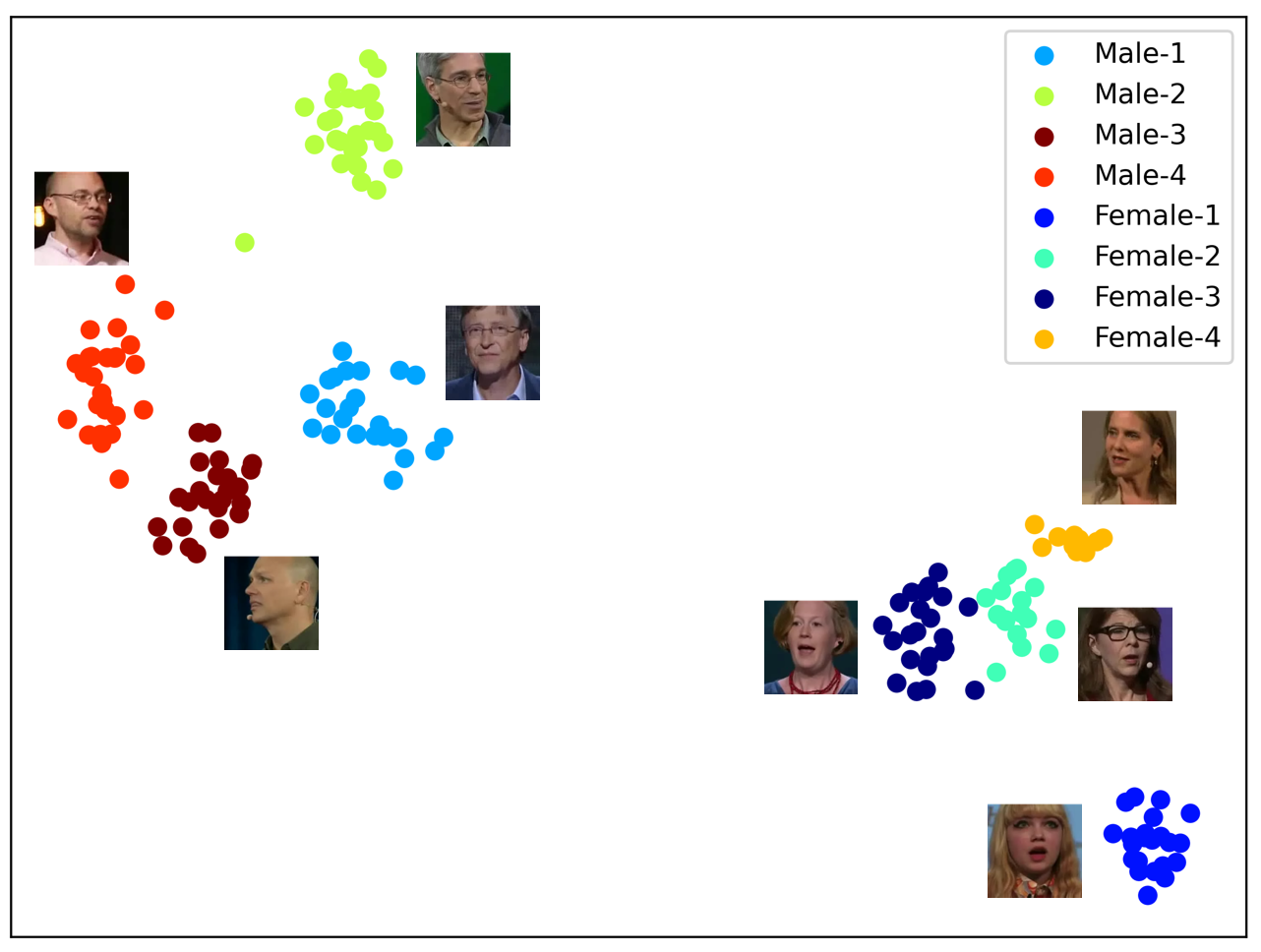}
    \vspace{-0.8em}
    \caption{The t-SNE visualization of speaker embeddings form generated speech. Each point represents a voice sample, with nearby images showing the faces of the speakers.}
    \label{fig:tsne_speaker}
    \vspace{-1.2em}
\end{figure}

\subsection{Evaluation Metrics}
\subsubsection{Subjective Metrics.} We evaluate the Mean Opinion Score (MOS) for speech naturalness (nMOS) and speaker similarity (sMOS) with ratings from 8 listeners. Ratings are assigned using a five-point scale: 1=Bad, 2=Poor, 3=Fair, 4=Good, 5=Excellent. Additionally, we conduct two preference tests to assess the alignment between generated speech and facial images: (1) selecting the generated speech that best matches a given face, and (2) selecting the facial image that best matches the style of a given generated speech.

\subsubsection{Objective Metrics.} We utilize the UTMOS \cite{saeki2022utmos} to evaluate the overall quality of the generated speech, serving as an objective alternative to the nMOS. The robustness and content consistency of the generated speech are quantified using the word error rate (WER) and character error rate (CER), which are calculated using the Whisper \cite{radford2023robust}. Additionally, speaker embeddings extracted via Resemblyzer are used to compute the speaker encoder cosine similarity (SECS) between the generated speech and the true style speech of the same speaker. Due to the absence of the paired speech data, SECS does not ensure content consistency in a pair, but the relative SECS scores across models indicate the ability to accurately map facial images to speaker styles. Moreover, we have developed metrics for speaker embedding consistency (SEC), which measure the uniformity of speech outputs from different facial angles of the same speaker, and speaker embedding diversity (SED), which assess the variability among speech outputs from different speakers.

\subsection{Quantitative Result and Analysis}
\subsubsection{Comparison with SOTA Methods.}
We evaluate our method against four recent face-to-speech generation methods, as outlined in Table \ref{SOTA_compare}. Among these, FaceVC \cite{lu2021face}, SP-FaceVC \cite{weng2023zero}, and FVMVC \cite{sheng2023face} control speech content using audio from the source speaker, while FaceTTS \cite{lee2023imaginary} uses text as input. Our approach exhibits a notable improvement on the UTMOS metric, indicating enhanced audio quality. Notably, our method achieves lower WER and CER, benefiting from the efficient content refinement implemented by MIDD. Although our method slightly trails SP-FaceVC in terms of the SECS metric, it surpasses all other methods evaluated. It is important to highlight that SP-FaceVC scores 0.912 on the SED metric, indicating a high level of timbral convergence among different speakers, which suggests a tendency toward generating an ``average'' timbre. In contrast, our method demonstrates superior performance on the SED metric, effectively capturing the most task-relevant facial features. Additionally, our results on the SEC metric highlight the robustness of our method to variations in facial embeddings.

\begin{table}[t]
\centering
\caption{Results of ablation studies on different model components. Best performances are highlighted in bold, while second-best performances are underlined.}
\vspace{-0.9em}
\label{ablation}
\resizebox{0.49\textwidth}{!}{
    \begin{tabular}{ccccccccc}
    \toprule
    IAQ-CL &MIDD& $L_{id-f}$ & $L_{id-s}$ & UTMOS $\uparrow$ &WER $\downarrow$& CER $\downarrow$& SECS $\uparrow$\\
    \midrule
    & & & &2.945& 15.29\% & 10.04\% &0.693 \\
    \small \Checkmark&  & & &3.227  & 18.04\%&11.57\%&\textbf{0.726}\\ 
    \small \Checkmark& \small \Checkmark & &&\underline{3.266}&12.70\% & 8.63\% & 0.709\\
    \small \Checkmark& \small \Checkmark & \small \Checkmark & &3.236 &12.25\%&\underline{7.97\%}	&0.709 \\
    \small \Checkmark& \small \Checkmark & &\small \Checkmark&  3.221& \textbf{12.01\%} &8.01\%	&0.712\\
    \small \Checkmark& \small \Checkmark& \small \Checkmark&\small  \Checkmark&\textbf{3.286}& \underline{12.11\%}&\textbf{7.86\%} &\underline{0.713} \\
    \bottomrule
    \end{tabular}}
\vspace{-1.2em}
\end{table}

\subsubsection{Ablation Studies.}
We investigate the impact of different model components on ID-FaceVC by conducting the following ablation studies: (1) w/o IAQ-CL: Randomly selects a facial frame from a video and uses FaceNet to generate a 512-dimensional vector, which is then fused through self-attention mechanisms and linear layers. (2) w/o MIDD: Directly extracts speaker embeddings by the Resemblyzer and maps these features to face embeddings using self-attention mechanisms and linear layers. (3) w/o $L_{id-f}$ and (4) w/o $L_{id-s}$: Omits the corresponding loss function. Experimental results are shown in Table \ref{ablation}.

In contrast to using static encoders for direct facial feature extraction, the IAQ-CL module significantly improves voice generation quality and face-to-voice mapping by effectively capturing facial features relevant to speaking styles. The MIDD module efficiently purifies the extracted content information, enhancing the clarity of the generated speech. Although there is a slight reduction in the SECS, this likely results from a trade-off with some intonation-related style features while preserving semantic content. This focus on content plays a crucial role in the clear expression of voice content. Additionally, supervision based on both facial and voice characteristics of speakers further strengthens the model's ability to distinguish critical features, thus improving generalization across different speakers.

% \begin{figure}[t]
%     \vspace{-0.5em}
%     \centering
%     \begin{subfigure}[b]{0.77\columnwidth}
%         \centering
%         \includegraphics[width=\textwidth]{sup/angle_M.png}
%         \vspace{-1.6em}
%         \caption{}
%         \vspace{-0.2em}
%     \end{subfigure}
%     \hfill
%     \begin{subfigure}[b]{0.77\columnwidth}
%         \centering
%         \includegraphics[width=\textwidth]{sup/angle_F.png}
%         \vspace{-1.6em}
%         \caption{}
%         \vspace{-0.8em}
%     \end{subfigure}
%     \caption{Mel-spectrograms of speech generated from facial images of two target speakers, captured at different angles and against various backgrounds.}
%     \label{fig:faceangle}
%     \vspace{-1.5em}
% \end{figure}

\subsection{Qualitative Result and Analysis}
\subsubsection{Visualization of Controllable Speech.}
For text-based input, we visualize the Mel spectrograms under various emotional states and speaking speeds, as depicted in Figure \ref{fig:melcontrol}. In the ``whispering'' state, the generated audio exhibits a more dispersed energy pattern with an increase in high-frequency components due to the incomplete vibration of vocal cords typical in whispering. In contrast, in the ``angry'' state, the speaker's voice shows greater fluctuations and intensity, with a quicker frequency and broader dynamic range. As the speaking speed increases, the spectral energy distribution becomes more compact, reducing the intervals between syllables. Conversely, when the speed decreases, the energy distribution expands, and syllables lengthen. These observations demonstrate that ID-FaceVC performs well in controlling different emotions and speaking speeds.

\begin{figure}[t]
    \vspace{-0.2em}
    \centering
    \includegraphics[width=0.46\textwidth]{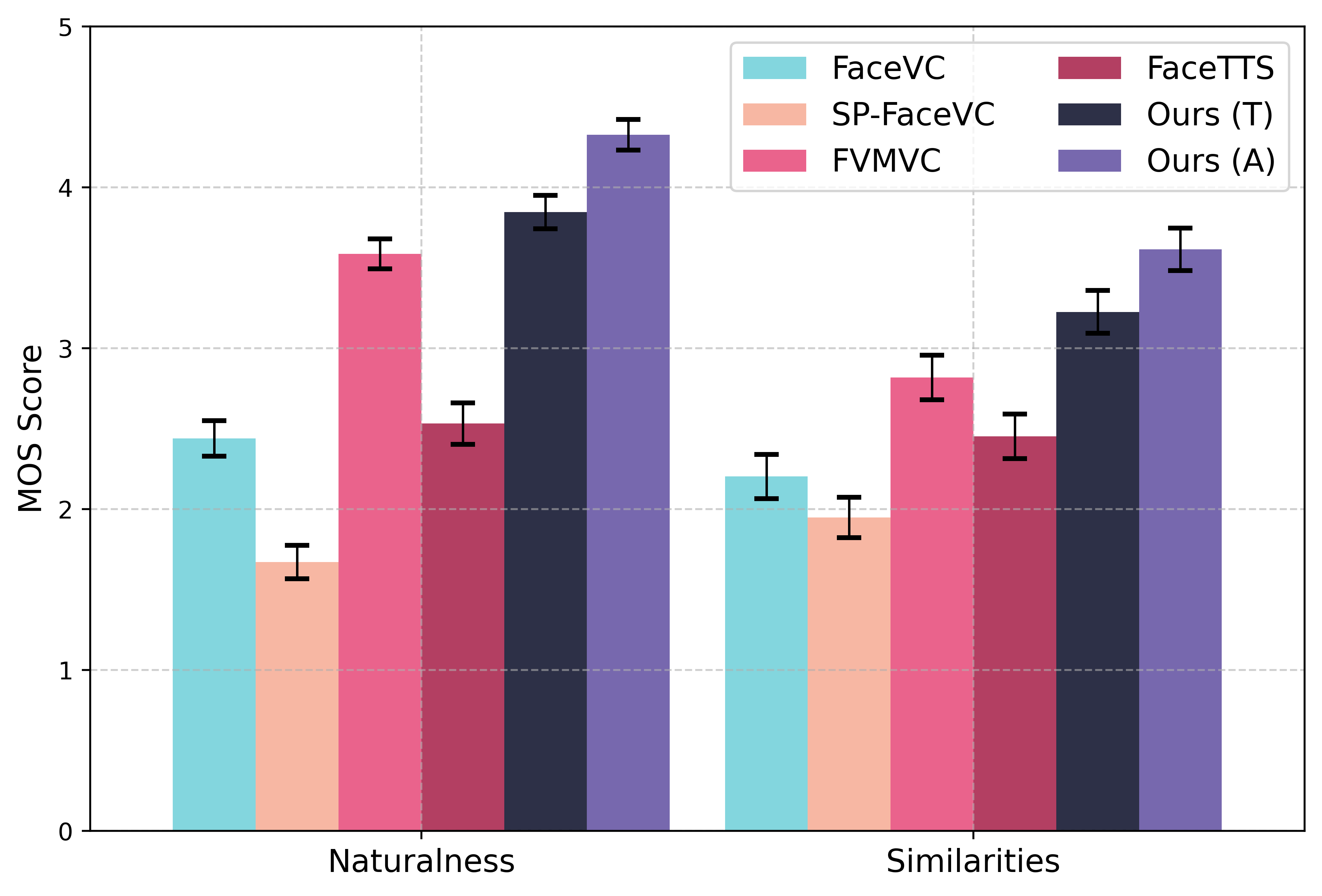}
    \vspace{-0.8em}
    \caption{User study results for naturalness and similarity metrics. Ours (T) and Ours (A) represent ID-FaceVC with text and audio inputs, respectively.}
    \label{fig:MOS}
    \vspace{-1.4em}
\end{figure}

\subsubsection{Style Manipulation.} We interpolate facial embeddings from two different speakers to generate various voice outputs, as shown in Figure \ref{fig:mix_faceemb}. As the facial embeddings transition from female to male, the fundamental frequency of the generated voice gradually decreases, and the harmonic distribution becomes denser. The voices in the intermediate transition phase not only retain high-frequency harmonic features typical of female voices but also incorporate low-frequency characteristics of male voices, illustrating a smooth transition in voice characteristics from female to male. This demonstrates our model's ability to precisely control voice output based on varying facial features.

\subsubsection{Distribution of Speaker Embedding.} For the generated speech, we use the Resemblyzer to extract speaker embeddings and visualize them using t-SNE, as depicted in Figure \ref{fig:tsne_speaker}. Voice samples generated from the same facial image form tight clusters, indicating that our model successfully maps unique vocal styles to different faces. Notably, embeddings for speakers of different genders display distinct distributions, with those of the same gender and similar ages showing closely matched speaker embeddings. This demonstrates our model's capability to effectively capture the most speech-relevant features from facial images.

\subsubsection{Visualization of Different Face Angles.}
We randomly selected two speakers and three facial images of each, captured from various angles, to perform ZS-FVC, as shown in Appendix Figure 9. Regardless of the facial expressions and angles, the voices generated by the model remained consistent across different images of the same speaker. This consistency is attributed to the model's ability to effectively align identity-related features in the faces with style-related features in the voice, demonstrating robustness to camera positions, backgrounds, and other noise.

\begin{figure}[t]
    \vspace{-0.1em}
    \centering
    \includegraphics[width=0.36\textwidth]{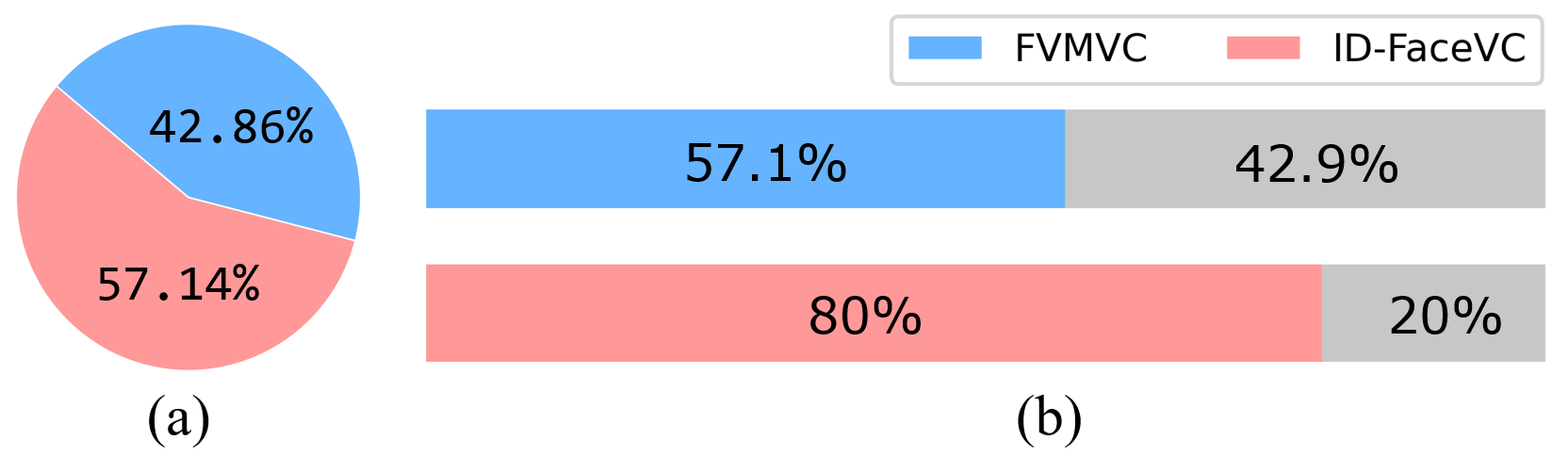}
    \vspace{-0.8em}
    \caption{Results of the preference test. (a) Given a facial image, preference results for choosing the more matching speech between outputs from FVMVC and ID-FaceVC. (b) Accuracy of selecting the more matching facial image given the model-generated speech.}
    \label{fig:preference}
    \vspace{-1.5em}
\end{figure}

 \vspace{-0.1em}
\subsection{User Study}
We evaluated the naturalness and similarity of generated speech through a user study involving eight experts. Each expert rated 27 sets of audio samples, with each set containing six comparative groups. As depicted in Figure \ref{fig:MOS}, our method consistently outperforms current SOTA approaches in both naturalness and similarity, while also exhibiting smaller confidence intervals. These findings demonstrate that ID-FaceVC reliably produces high-quality outputs with enhanced stability.

We further validate our model's ability to map facial features to speech characteristics through preference tests. To increase the challenge of the experiment, we conduct gender-matched tests, selecting face and audio samples from individuals of the same gender. As depicted in Figure \ref{fig:preference}, in the face-based preference test, 57.14\% of evaluators believe that ID-FaceVC produces results that better match the given facial images. In the voice-based preference test, evaluators correctly identify the match 22.9\% more often when the speech is generated by ID-FaceVC rather than FVMVC, demonstrating that the speech generated by ID-FaceVC more accurately aligns with the corresponding facial images.

\vspace{-0.3em}
\section{Conclusion}
\label{sec:Conclusion}
In this work, we introduce a novel ID-FaceVC framework, effectively generating speech that aligns with facial identity features. Our framework includes the IAQ-CL and MIDD modules to precisely map facial features to speech. Additionally, we incorporate text as an alternative modality for controlling speech content and employ a style controllable strategy that ensures speech generated from text is natural, rhythmic, and controllable. Both quantitative and qualitative experiments validate the overall effectiveness of our framework and the individual modules. Future work aims to expand beyond audio generation to include expressive facial animations, transitioning from merely ``audible" to ``both audible and visible."

% \subsubsection{Acknowledgments.}
% The acknowledgments section, if included, appears after the main body of text and is headed ``Acknowledgments." This section includes acknowledgments of help from associates and colleagues, credits to sponsoring agencies, financial support, and permission to publish. Please acknowledge other contributors, grant support, and so forth, in this section. Do not put acknowledgments in a footnote on the first page. If your grant agency requires acknowledgment of the grant on page 1, limit the footnote to the required statement, and put the remaining acknowledgments at the back. Please try to limit acknowledgments to no more than three sentences. 

% \subsubsection{Appendices.}
% Any appendices follow the acknowledgments, if included, or after the main body of text if no acknowledgments appear. 

% \bibliographystyle{aaai25}
\bibliography{ref}

\clearpage
\section{Appendix}

\subsection{Visualization of Different Face Angles.}
As shown in Appendix Figure \ref{fig:faceangle}, the speech styles generated by ID-FaceVC from different facial images of the same target speaker show consistency.
\begin{figure}[h]
    \vspace{-0.5em}
    \centering
    \begin{subfigure}[b]{0.77\columnwidth}
        \centering
        \includegraphics[width=\textwidth]{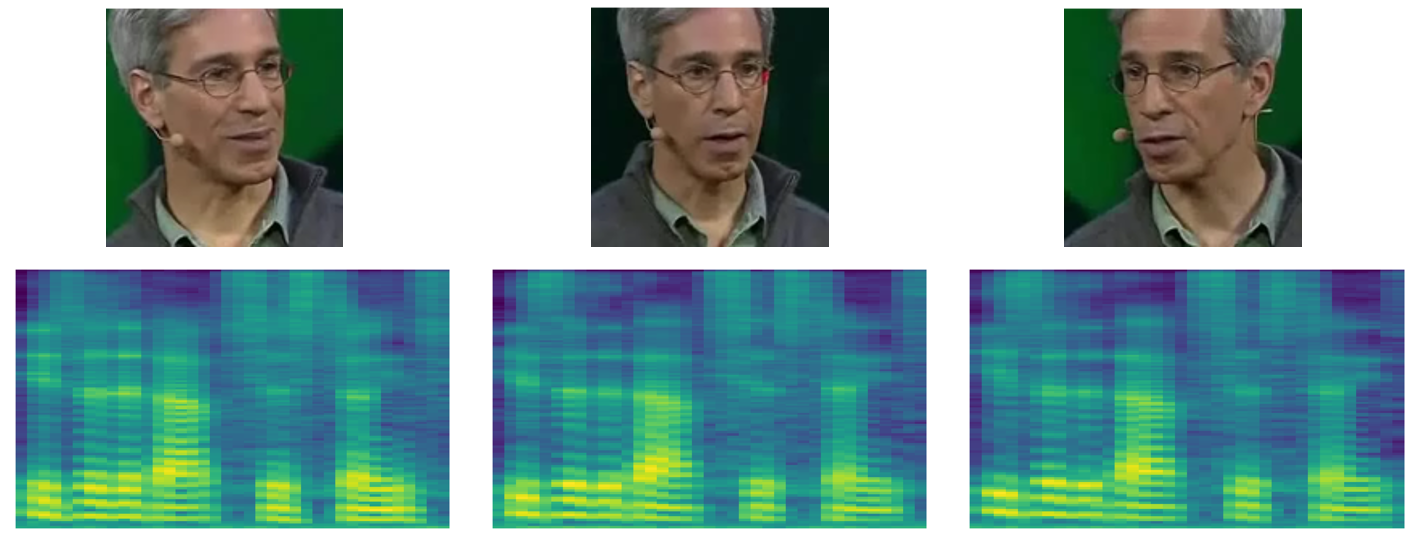}
        \vspace{-1.6em}
        \caption{}
        \vspace{-0.2em}
    \end{subfigure}
    \hfill
    \begin{subfigure}[b]{0.77\columnwidth}
        \centering
        \includegraphics[width=\textwidth]{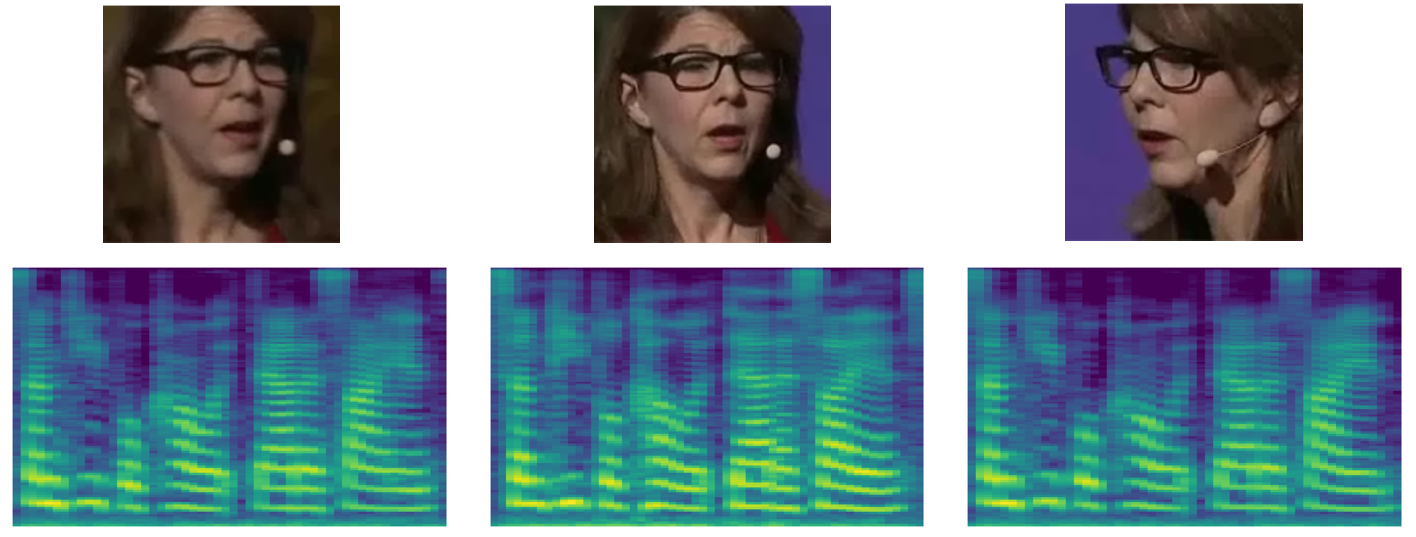}
        \vspace{-1.6em}
        \caption{}
        \vspace{-0.8em}
    \end{subfigure}
    \caption{Mel-spectrograms of speech generated from facial images of two target speakers, captured at different angles and against various backgrounds. }
    \label{fig:faceangle}
    \vspace{-1.5em}
\end{figure}

\end{document}